\documentclass[a4paper,11pt]{article}
\usepackage{amsmath,amssymb,revsymb,braket,feynmp,subfigure}
\usepackage[dvipdfmx]{graphicx,color}
\usepackage[title,titletoc,toc]{appendix}
\usepackage{authblk,setspace}
\usepackage[bf,labelsep=period,figurename=FIG.\ , tablename=TABLE]{caption} 
\numberwithin{equation}{section}
\usepackage{amsmath}	
\usepackage{graphicx}	

\begin{document}
\begin{titlepage}
\begin{flushright}
 KEK-TH-1845
\end{flushright}
\begin{center}
\begin{spacing}{2.5}
 {\LARGE Lepton flavor violating $Z$-boson couplings from non-standard Higgs interactions}\\
\end{spacing}
\medskip
{\large Toru Goto$^1$\footnote{tgoto@post.kek.jp}, Ryuichiro Kitano$^{1,2}$\footnote{Ryuichiro.Kitano@kek.jp}, and Shingo Mori$^2$\footnote{smori@post.kek.jp}}\\
\medskip
${}^1${\it KEK Theory Center, Tsukuba 305-0801, Japan}\\
${}^2${\it Department of Particle and Nuclear Physics,\par The Graduate University for Advanced Studies (Sokendai),\par Tsukuba 305-0801, Japan}\\
\medskip
{\today}
\end{center}
\medskip
\begin{abstract}
The Standard Model predicts that the Higgs boson couples to the fermions in the mass eigenstates.
We consider the effects of lepton flavor violating (LFV) $Z$ boson couplings in the case where the Higgs boson has flavor non-diagonal Yukawa interactions with the muon and the tau lepton generated from physics beyond the Standard Model.
We list the formulae of the couplings of the effective interactions among the $\tau$ lepton, the muon and the $Z$ boson.
 Using these formulae, we calculate the branching fractions of various leptonic and hadronic LFV $\tau$ decays, and the LFV $Z$ boson decay: $Z\to \tau\mu$. 
Although the $Z$-boson contributions to LFV tau decays cannot be ignored in terms of the counting of operator dimensions or chirality flipptings, it turns out that they are not very significant for $\tau\to3\mu$ and $\tau\to\mu\rho$ decays.
We also calculate the branching fractions of the processes, $\tau\to \mu\pi$, $\tau\to \mu\eta^{(\prime)}$ and $\tau\to\mu a_1$, which are dominated by the $Z$-boson exchanges due to the spin and the parity of the hadrons.
\end{abstract}
\end{titlepage}
\newpage
\section{Introduction}

The Higgs boson has been discovered at LHC~\cite{Aad:2012tfa,Chatrchyan:2012ufa} in 2012 with mass around $125$ GeV~\cite{Aad:2013wqa,Chatrchyan:2012jja}, and all particles in the Standard Model (SM) are now discovered.
Nevertheless, the nature of the Higgs field and its potential is still unknown. 
It is quite natural to expect that there is some larger framework, physics beyond the standard model (BSM), that provides us with better understanding of the nature of the Higgs boson.

The lepton flavor violation (LFV) is one of the clear signals of BSM.
Although the observation of neutrino oscillation phenomena (see Ref.~\cite{Mohapatra:2005wg} and references therein) implies that lepton flavor is not conserved in the neutrino sector, simple incorporation of the neutrino mass in the SM does not result in the charged lepton flavor violation at the observable level~\cite{Petcov:1976ff,Lee:1977tib}.
Since the Higgs field is the origin of the flavor structure in the SM, it is quite conceivable that BSM hidden behind the Higgs mechanism induces the LFV processes with their rates much larger than the ones predicted from the neutrino mixings.
Indeed, it has been shown that various scenarios of BSM predict large branching ratios of LFV processes, for example in supersymmetric models~\cite{Borzumati:1986qx,Barbieri:1994pv,Barbieri:1995tw,Hisano:1995cp,Hisano:1995nq,Hisano:1996qq}, multi-Higgs doublet models~\cite{Bjorken:1977vt}, the littlest Higgs model with T-parity~\cite{Goto:2010sn,Blanke:2009am,delAguila:2008zu}, and the SM extended by extra dimension~\cite{Kitano:2000wr,Grossman:1999ra}.

After the recent discovery of the Higgs boson, there have been many studies on its properties such as spin-parity and couplings~\cite{Aad:2013wqa,Chatrchyan:2012jja}.
One of the non-trivial predictions of the SM is that the Higgs boson couples to fermions in the mass eigenstates~\cite{Glashow:1976nt}.
This prediction is not necessarily true in BSM physics, and thus searching for flavor off-diagonal couplings opens up the opportunities to probe the physics behind the Higgs mechanism.
The possibility of LFV couplings of the Higgs boson has been originally studied by Bjorken and Weinberg~\cite{Bjorken:1977vt} (see also~\cite{Shanker:1981mj,Pilaftsis:1992st}), and after the start of the LHC experiments the model has been paid a renewed attention~\cite{Blankenburg:2012ex,Harnik:2012pb,Celis:2013xja,Omura:2015nja}. 
In these works LFV processes induced by the off-diagonal Higgs couplings have been discussed, especially $\mu\to e\gamma$, $\tau\to \mu \gamma$ and $\tau\to e\gamma$ decays.
Bjorken and Weinberg have pointed out that two-loop diagrams provide the main contribution to these processes rather than one-loop diagrams due to the chirality structure.
It has later been estimated that the two-loop contributions are a factor of a few larger than the one-loop ones~\cite{Harnik:2012pb,Chang:1993kw}.

LFV processes have been searched for in the decays of the muon and the tau lepton.
The current experimental limits on the branching fractions are ${\cal B}(\mu\to e\gamma)<5.7\times10^{-13}$~\cite{Adam:2013mnn}, ${\cal B}(\mu\to3e)<1.0\times10^{-12}$~\cite{Bellgardt:1987du} and ${\cal B}(\mu {\rm Au}\rightarrow e {\rm Au})<7\times10^{-13}$~\cite{Bertl:2006up} for muons.
For LFV tau decays, there have been searches for decay modes such as $\tau\to\mu\gamma$, $\tau\to 3\ {\rm leptons}$, and $\tau\to\mu+{\rm hadrons}$~\cite{Lusiani:2010eg}, and the upper bounds have been obtained as ${\cal O}(10^{-8})$. 
In particular, the BaBar and Belle experiments set a limit ${\cal B}(\tau\to\mu\gamma)<4.4\times10^{-8}$~\cite{Aubert:2009ag,Hayasaka:2007vc}. 
Although the upper bounds look much stronger for muon decays, tau decays may be more important if the origin of LFV is related to the Higgs mechanism as the Higgs field couples strongly to matter in the third generation.
In future, the sensitivity to the branching ratios of the LFV tau decays such as $\tau\rightarrow\mu\gamma$ and $\tau\to3\mu$ will be improved to at the level of $10^{-(9-10)}$~\cite{FPCP_Barrett:2015} at the Belle~II experiment.

In addition to the B factory experiments, there have been searches for the LFV Higgs decays at the LHC experiments.
Recently, the CMS collaboration has reported the upper limit of the LFV Higgs decay into the tau and the muon using $19.7\ {\rm fb}^{-1}$ of $\sqrt{s}=8\ {\rm TeV}$ data set taken in 2012~\cite{Khachatryan:2015kon}, and this provides us with the upper limit on LFV Yukawa couplings of ${\cal O}(10^{-2})$~\cite{Harnik:2012pb,Khachatryan:2015kon}.
Interestingly, the analysis of CMS has also reported a $2.4$ $\sigma$ excess in the $H\rightarrow\tau\mu$ decay mode~\cite{Khachatryan:2015kon}. 

In the literature on the LFV tau decays through the non-diagonal Yukawa couplings, one-loop and two-loop photon mediated diagrams and the tree-level Higgs exchange diagrams are evaluated for $\tau\to \mu \gamma$, $\tau\to3\mu$ and $\tau\to \mu +{\rm hadron}$ decay processes~\cite{Harnik:2012pb,Celis:2013xja}.
According to their formulae the sensitivities to the LFV couplings at the future B factory experiments are found to be similar to those at the LHC experiments.

In this paper, we discuss the $Z$ boson mediated contributions to the LFV tau decays in the model with the flavor off-diagonal Higgs interactions.
Such contributions have not been calculated in literature although, based on the dimensional analysis and the chirality structure, the contributions of the $Z$ boson to effective LFV four-fermion operators are expected to be comparable to other diagrams, such as photon mediated diagrams.
Moreover, in $\tau$ decays there are final states which are allowed only through the $Z$ boson.
We first list the coefficients of effective $\tau\mathchar`-\mu\mathchar`-Z$ interactions for both the monopole and the dipole types.
Although the dimension of the dipole operator is higher than that of the monopole type, one cannot ignore the dipole one since the contribution to the decay amplitude is comparable considering the chirality structure.
Using these coefficients, we evaluate the LFV tau decays as well as LFV $Z$ decays.
It turns out that the $Z$ boson mediated contributions to the tau decays are numerically sub-dominant when there are contribution from photon mediated diagrams such as $\tau\to 3\mu$ and $\tau\to \mu \rho$ decays. 
For the hadronic decays into pseudoscalar and axial vector mesons, the $Z$ boson contributions are dominant, and the dipole operators are found to be numerically less important compared to the monopole ones.

In Sec.~\ref{sec:1LFVH}, we discuss the model of the LFV Yukawa couplings and gauge invariant higher dimensional operators which generate the couplings. 
In Sec.~\ref{sec:2taumuZ} we list the coefficients of the effective $\tau\mathchar`-\mu\mathchar`-Z$ interaction obtained by evaluating the one-loop $Z$ penguin diagrams, then by using the coefficients we evaluate the LFV $Z$ decay.
By using the effective Lagrangian, we calculate the branching ratios of LFV tau decays in Sec.~\ref{sec:4taudecay}. We summarize our result in Sec.~\ref{sec:5summary}.
\section{Lepton Flavor Violating Higgs couplings}\label{sec:1LFVH}
Although the couplings of the SM Higgs boson are predicted to be flavor diagonal, the BSM contribution represented by higher dimensional operators can modify the prediction.
In this section, we review how such flavor off-diagonal Yukawa couplings are generated from dimension-six operators.

One of such operators is $\bar{L}e_R\varphi (\varphi^\dagger \varphi)$, where $L$ and $\varphi$ are left-handed lepton doublets and the Higgs doublet, respectively, and $e_R$ is right-handed lepton fields.
The Lagrangian which is responsible for the Yukawa interaction of lepton fields is
\begin{gather}
{\cal L}_{\rm Yukawa}=-y_{ij}\bar{L}^ie_R^j\varphi-\frac{C_{ij}}{\Lambda^2}\bar{L}^ie_R^j\varphi(\varphi^\dagger \varphi)+{\rm H.c.},
\end{gather}
where $y_{ij}$ and $C_{ij}$ are dimensionless coupling constants.
Indices $i,j=1,2,3$ indicate the generation of the lepton fields, and $\Lambda$ is an energy scale for the normalization.
The Yukawa terms ${\cal L}_{\rm Yukawa}$ are expanded around the VEV of the Higgs boson as follows:
\begin{eqnarray}
 {\cal L}_{\rm Yukawa}&=&-\tilde{m}_{ij}\bar{e}_L^ie_R^j-\tilde{Y}_{ij}\bar{e}_L^ie_R^jH-i\tilde{Y}_{ij}\bar{e}_L^ie_R^j\phi_2+{\rm H.c.}\nonumber\\
&&-\left(\frac{v}{\sqrt{2}\Lambda^2}\right)C_{ij}\bar{e}_L^ie_R^j\left[\phi^-\phi^++\frac{1}{2}\left(3H^2+2iH\phi_2+\phi^2_2\right)\right]+{\rm H.c.}+\cdots,\label{eq:effYukawa}\\
\tilde{m}_{ij}&=&\frac{v}{\sqrt{2}}\left(y_{ij}+\left(\frac{v}{\sqrt{2}\Lambda}\right)^2C_{ij}\right),\label{eq:effY1}\\
\tilde{Y}_{ij}&=&\frac{1}{\sqrt{2}}\left(y_{ij}+\left(\frac{v}{\sqrt{2}\Lambda}\right)^23C_{ij}\right),\label{eq:effY2}
\end{eqnarray}
where in a general gauge the Higgs doublet is expanded around its vacuum expectation value (VEV), $v$, as $\varphi=(\phi^+,(v+H+i\phi_2)/\sqrt{2})^{\rm T}$.
The charged and the neutral Nambu-Goldstone bosons are denoted by $\phi^+$ and $\phi_2$, respectively.
One can see that the factor of three in Eq.~(\ref{eq:effY2}) prevents us from simultaneously diagonalizing the mass and the Yukawa matrices unless $C_{ij}=0$ or $C_{ij}\propto y_{ij}$.
In the mass eigenstate of lepton fields with mass matrix $m={\rm diag}(m_e,m_\mu,m_\tau)$, the Yukawa matrix $Y$ has, thus in general, non-vanishing off-diagonal elements.
In addition, there are new flavor violating interactions such as the fourth term in Eq.~(\ref{eq:effYukawa}), which is necessary to maintain gauge invariance.

In this work, we consider the effects of $\tau\mathchar`-\mu\mathchar`-H$ interaction, and we assume other off-diagonal elements are zero for simplicity.
In the mass basis of the charged leptons, the Lagrangian is given by
\begin{gather}
 {\cal L}_{\rm Yuakwa}=-m_{l}\bar{l}_Ll_R-Y_{l}\bar{l}_Ll_RH-\bar{\mu}\left(Y_{\mu\tau}P_R+Y_{\tau\mu}^*P_L\right)\tau H+{\rm H.c.}+\cdots,\qquad (l=e,\mu,\tau),
\end{gather}
where $P_R=(1+\gamma_5)/2$ and $P_L=(1-\gamma_5)/2$.
We also assume that the diagonal components of the Yukawa matrix is SM-like, i.e., $Y_l=m_l/v$.

These LFV Yukawa couplings are directly constrained by the searches for $H\to\tau\mu$ process at the LHC experiments.
The current experimental upper bound is ${\cal B}(H\to\tau^\pm\mu^\mp)={\cal B}(H\to\tau^+\mu^-)+{\cal B}(H\to\tau^-\mu^+)\leq1.51\times10^{-2}$~\cite{Khachatryan:2015kon}, whereas the theoretical prediction is
\begin{gather}
 {\cal B}(H\rightarrow\tau^\pm\mu^\mp) =1.2\times10^{3}\left(|Y_{\tau\mu}|^2+|Y_{\mu\tau}|^2\right),
\end{gather}
where the Higgs boson mass is taken as $m_H=125$ GeV and we use the SM prediction of its width $\Gamma_H=4.0$ MeV~\cite{Heinemeyer:2013tqa} at $m_H=125$ GeV in the evaluation.
\section{$\tau\mathchar`-\mu\mathchar`-Z$ from LFV Yukawa coupling}\label{sec:2taumuZ}
\subsection{Calculation of $\tau\mathchar`-\mu\mathchar`-Z$ vertex from one-loop amplitudes}
In the presence of the flavor violating Yukawa interactions, the effective $\tau\mathchar`-\mu\mathchar`-Z$ couplings are induced by one-loop diagrams.
The $\tau\to\mu Z^*$ transition amplitude is in general parametrized as follows:
\begin{eqnarray}
{\cal M}(\tau^-\to\mu^- Z^{0*}(q))&=&
-\bar{u}_\mu\left(\frac{B_R^{Z*}(s)}{m_\tau}P_R+\frac{B_L^{Z*}(s)}{m_\tau}P_L\right)u_\tau iq^\mu\epsilon_\mu^*(q)\nonumber\\
&&-\bar{u}_\mu \gamma^\mu \left(C_R^{Z*}(s)P_R+C_L^{Z*}(s)P_L\right)u_\tau\epsilon_\mu^*(q)\nonumber\\
&&-\bar{u}_\mu \sigma^{\mu\nu}\left(\frac{D_R^{Z*}(s)}{m_\tau}P_R+\frac{D_L^{Z*}(s)}{m_\tau}P_L\right)u_\tau 2iq_\mu \epsilon_\nu^*(q),\label{eq:Mz}
\end{eqnarray}
where $q^\mu$ is the four momentum of the off-shell $Z$ boson, and $s=q^2$ and $\sigma^{\mu\nu}=(i/2)[\gamma^\mu,\gamma^\nu]$.
The wave functions of the muon and the tau lepton are denoted by $u_\mu$ and $u_\tau$, respectively.
The polarization vector of the $Z$ boson is denoted by $\epsilon^\mu(q)$.
We set the initial $\tau$ and $\mu$ in the final state as on-shell.

The dimensionless effective couplings $B_{R,L}^Z(s)$, $C_{R,L}^Z(s)$ and $D_{R,L}^Z(s)$ are in general functions of $s$.
In the calculation of $Z$ boson decays and tau decays, the contribution from $B_{R,L}^Z(s)$ can be neglected since $q_\mu \epsilon^\mu=0$ for an on-shell $Z$ boson and $q_\mu J^\mu={\cal O}(m_f)$ where $J^\mu$ and $m_f$ are respectively the current and the mass of the light fermions in the final states.
Although the terms with couplings $C_{R,L}^Z(s)$ and $D_{R,L}^Z(s)$ respectively correspond to dimension-four and five operators, the ones with $D_{R,L}^Z(s)$ can be equally important in general as we will see later.

\begin{figure}[t]
\begin{center}
 \subfigure[]{\includegraphics[width=50mm]{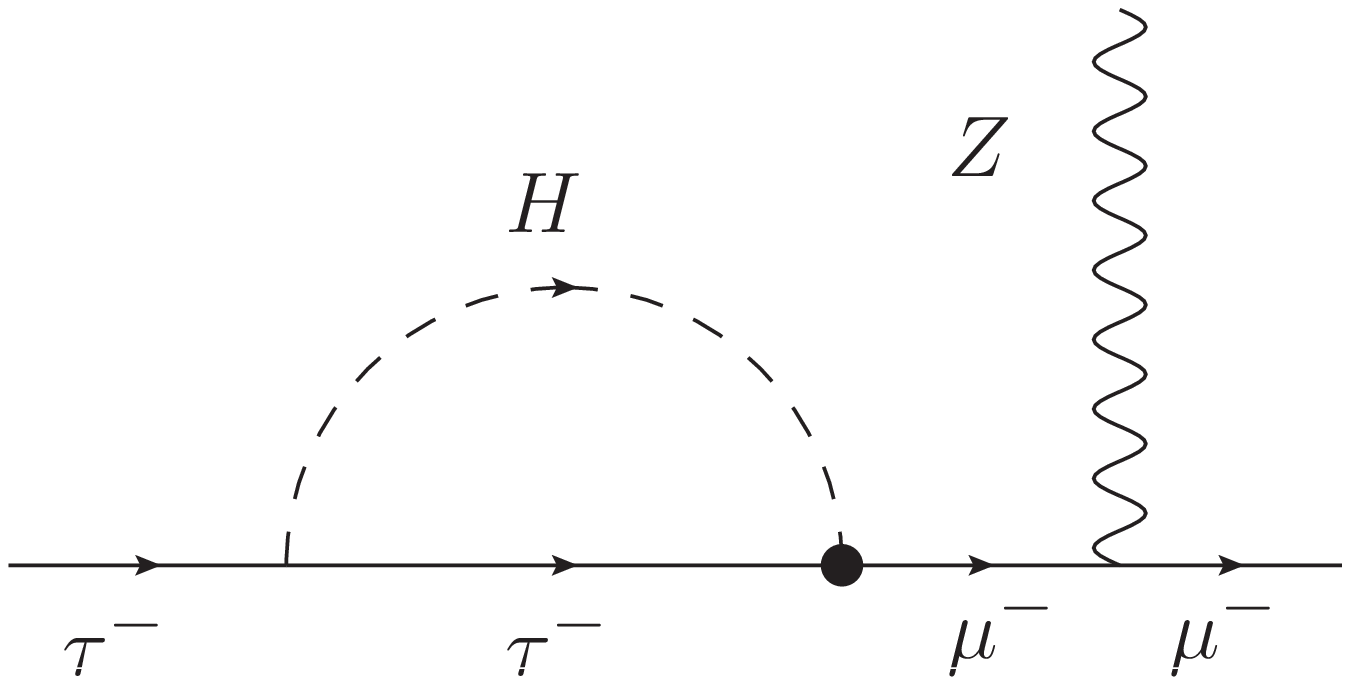}\label{fig:tau_muZa}}
 \subfigure[]{\includegraphics[width=50mm]{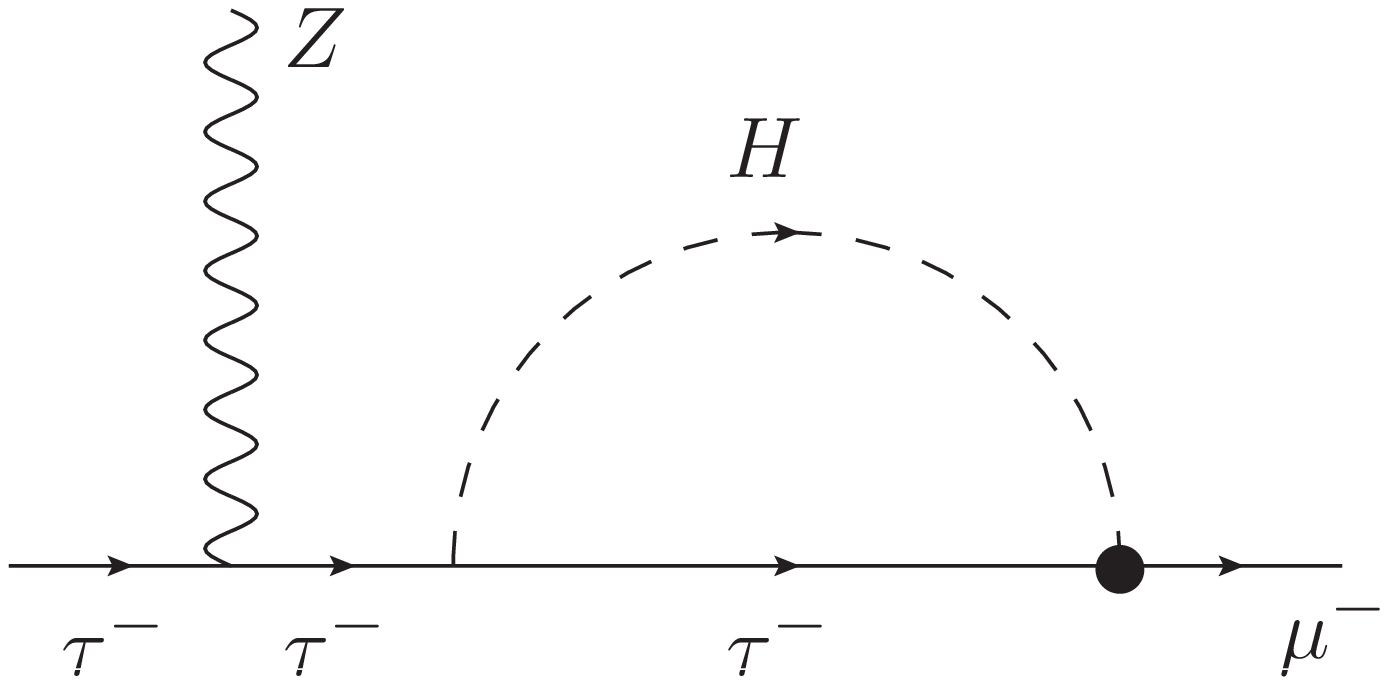}\label{fig:tau_muZb}}
 \subfigure[]{\includegraphics[width=50mm]{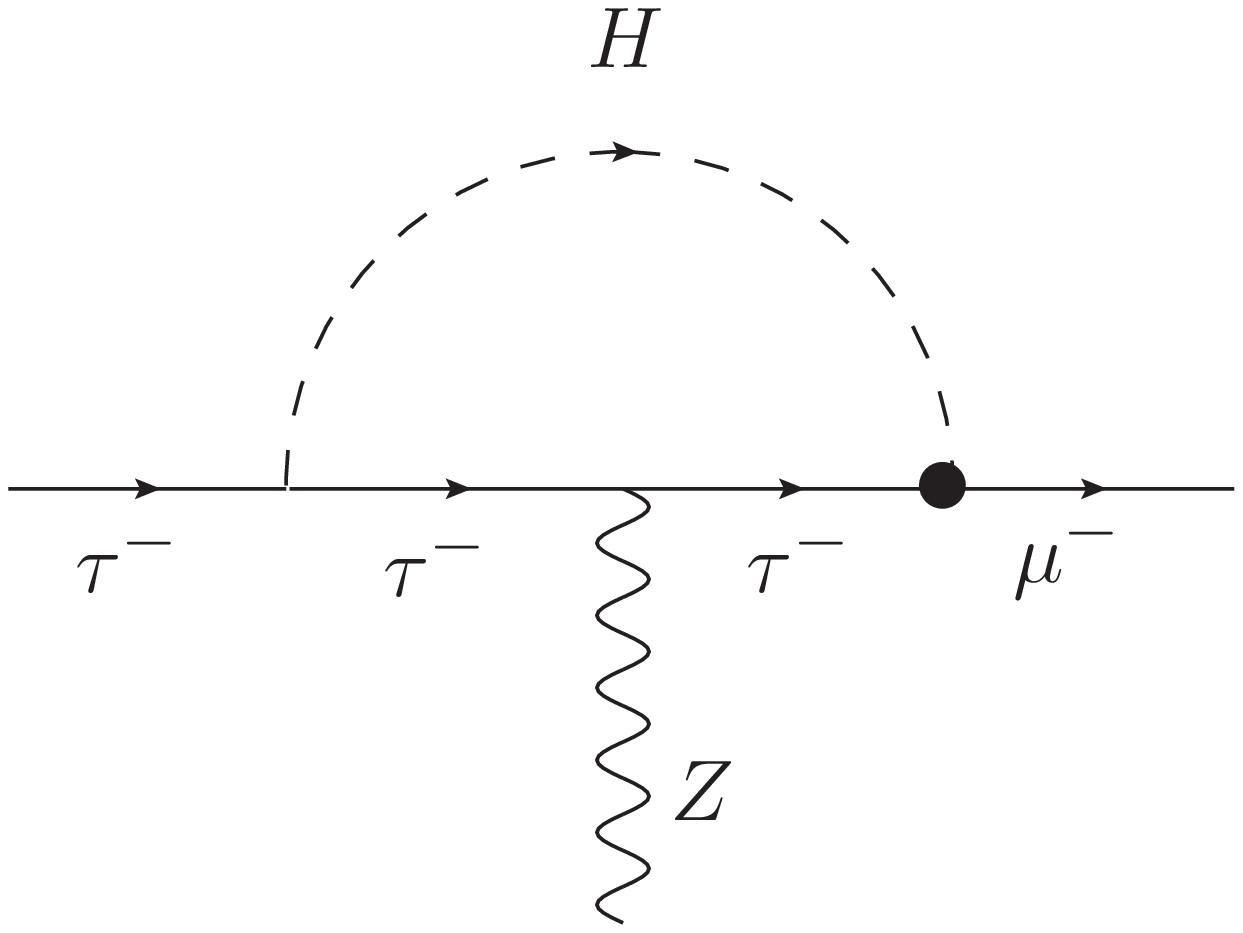}\label{fig:tau_muZc}}\\
 \subfigure[]{\includegraphics[width=50mm]{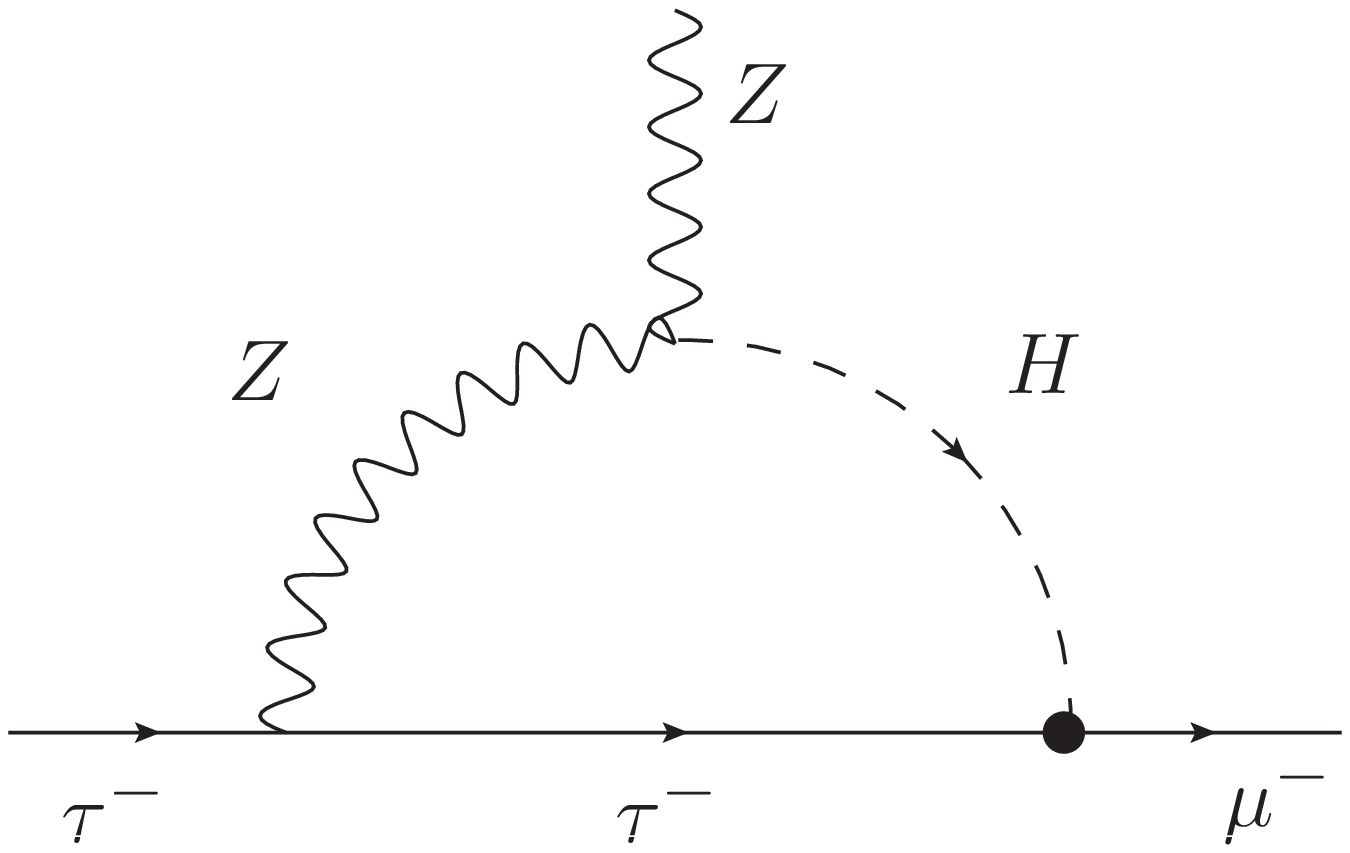}\label{fig:tau_muZd}}
 \subfigure[]{\includegraphics[width=50mm]{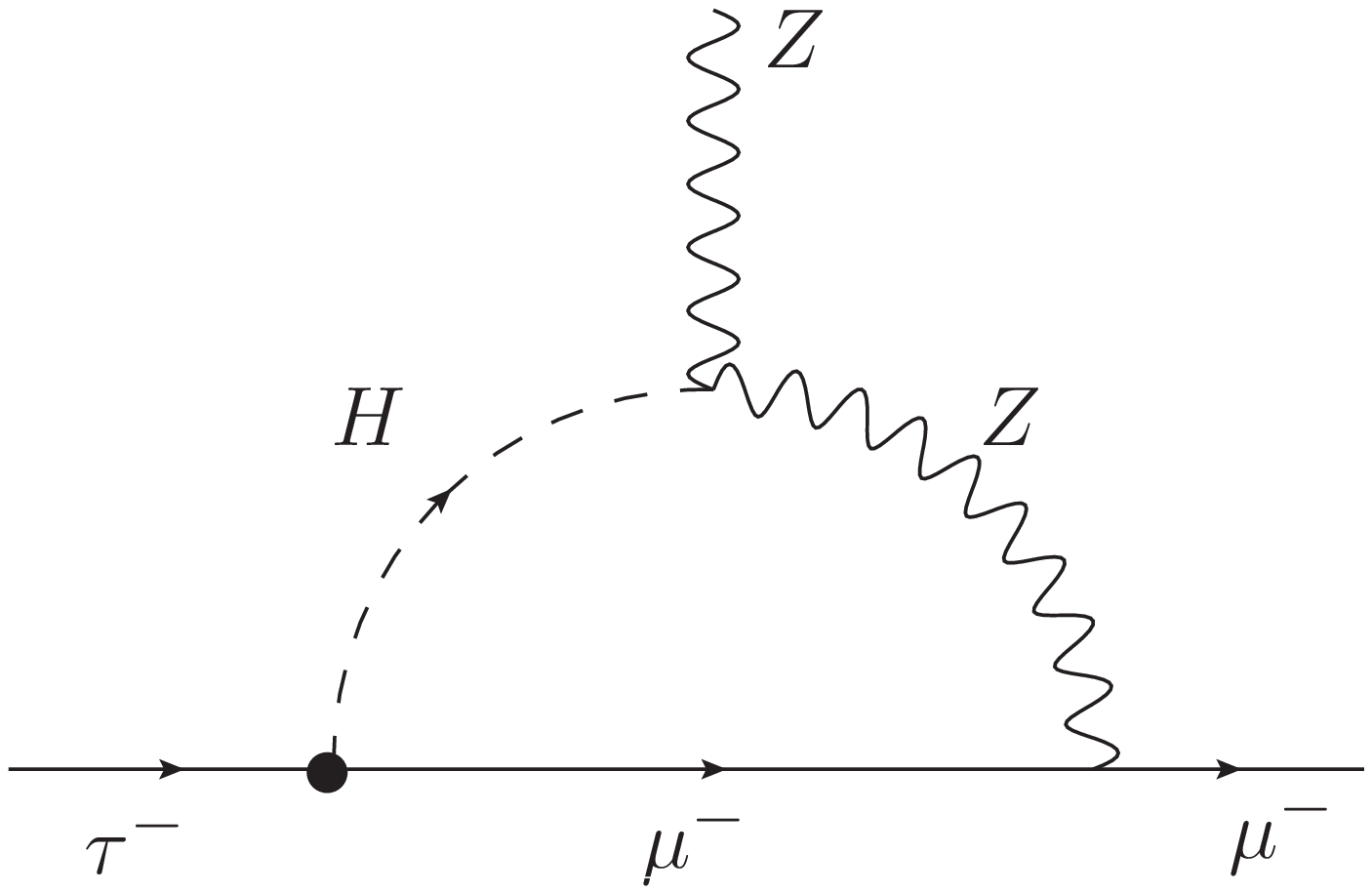}\label{fig:tau_muZe}}
 \subfigure[]{\includegraphics[width=50mm]{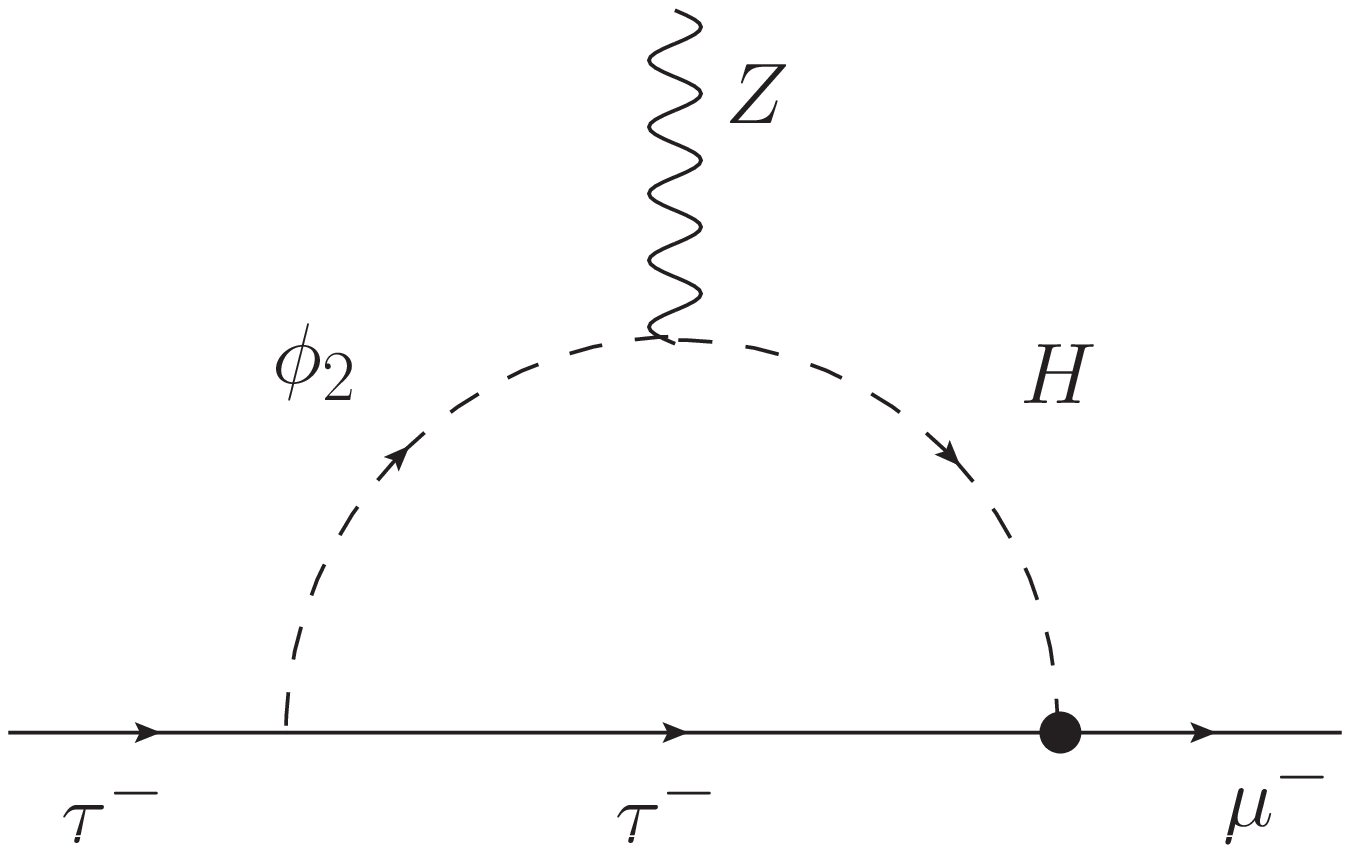}\label{fig:tau_muZf}}\\
\caption{One-loop finite and gauge invariant set of diagrams which contribute to $\tau\mathchar`-\mu\mathchar`-Z$ vertex.  Black dot denotes the LFV Yukawa coupling.}\label{fig:tau_muZ}
\end{center}
\end{figure}

The one-loop level diagrams which generate the above amplitudes are shown in FIG.~\ref{fig:tau_muZ}, where we omit diagrams which are suppressed by the muon Yukawa coupling.
Summing all six diagrams we obtain a finite result which does not depend on a gauge parameter.

By evaluating diagrams in Fig.\ref{fig:tau_muZ}, we obtain
\begin{eqnarray}
 C_{L}^Z(s)&=&\frac{g_Z Y_\tau Y_{\tau\mu}}{64\pi^2}\left[F_V^v(s)g_V^e+F_V^a(s)g_{A}^e\right],\\
 C_{R}^Z(s)&=&\frac{g_Z Y_\tau Y_{\mu\tau}^*}{64\pi^2}\left[F_V^v(s)g_V^e-F_V^a(s)g_{A}^e\right],\\
 D_{L}^Z(s)&=&\frac{g_Z Y_\tau Y_{\mu\tau}^*}{64\pi^2}\left[F_{D}^v(s)g_{V}^e+F_{D}^a(s)g_{A}^e\right],\label{eq:DLs}\\
 D_{R}^Z(s)&=&\frac{g_Z Y_\tau Y_{\tau\mu}}{64\pi^2}\left[F_{D}^v(s)g_{V}^e-F_{D}^a(s)g_{A}^e\right],\label{eq:DRs}
\end{eqnarray}
where the functions $F_V^{v,a}(s)$ and $F_{D}^{v,a}(s)$ are expressed in terms of the Passarino-Veltman functions $B_0$, and $C_i,\ (i=1,2,12,22,00)$ defined in App.~\ref{sec:PVfunction} as follows:
\begin{eqnarray}
 F_V^v(s)&=&(2m_\tau^2(C_1+C_2+C_{12}+C_{22})+4C_{00})(0,s,m_\tau^2,m_H^2,m_\tau^2,m_\tau^2)\nonumber\\
&&+2(2m_\tau^2-m_H^2)C_0(m_\tau^2,s,0,m_H^2,m_\tau^2,m_\tau^2)\nonumber\\
&&+4m_Z^2C_0(s,m_\tau^2,0,m_H^2,m_Z^2,m_\tau^2)+4m_Z^2(C_0+C_2)(s,0,m_\tau^2,m_H^2,m_Z^2,0)\nonumber\\
&&+\left(4-\frac{m_H^2}{m_\tau^2}\right)\left(B_0(m_\tau^2,m_H^2,m_\tau^2)-B_0(0,m_H^2,m_\tau^2)\right)\nonumber\\
&&+B_0(0,m_H^2,m_\tau^2)-2B_0(s,m_\tau^2,m_\tau^2),\\
F_V^a(s)&=&(2m_\tau^2(-C_1-C_2+C_{12}+C_{22})+4C_{00})(0,s,m_\tau^2,m_H^2,m_\tau^2,m_\tau^2)\nonumber\\
&&-2(m_H^2+2m_\tau^2 )C_0(m_\tau^2,s,0,m_H^2,m_\tau^2,m_\tau^2)-4m_Z^2(C_0+C_2)(s,0,m^2_\tau,m_H^2,m_Z^2,0)\nonumber\\
&&+4(-m_\tau^2 C_2+m_\tau^2 C_{12}-m_Z^2 C_0+2C_{00})(s,m_\tau^2,0,m_H^2,m_Z^2,m^2_\tau)\nonumber\\
&&+\left(\frac{m_H^2}{m_\tau^2}-4\right)\left(B_0(m_\tau^2,m_H^2,m_\tau^2)-B_0(0,m_H^2,m_\tau^2)\right)\nonumber\\
&&-B_0(0,m_H^2,m_\tau^2)-2B_0(s,m^2_\tau,m^2_\tau),
\end{eqnarray}
\begin{eqnarray}
  F_D^v(s)&=&-m_\tau^2(C_1+2C_2+C_{12}+C_{22})(0,s,m_\tau^2,m_H^2,m_\tau^2,m_\tau^2)\nonumber\\
&&+2m_Z^2\left(C_1(s,0,m_\tau^2,m_H^2,m_Z^2,0)+C_1(s,m_\tau^2,0,m_H^2,m_Z^2,m_\tau^2)\right),\\
 F_D^a(s)&=&-m_\tau^2(C_1-2C_2-C_{12}-C_{22})(0,s,m_\tau^2,m_H^2,m_\tau^2,m_\tau^2)\nonumber\\
&&-2m_\tau^2( C_2+C_{12})(s,m_\tau^2,0,m_H^2,m_Z^2,m_\tau^2)\nonumber\\
&&+2m_Z^2\left(C_1(s,0,m_\tau^2,m_H^2,m_Z^2,0)-C_1(s,m_\tau^2,0,m_H^2,m_Z^2,m_\tau^2)\right).\label{eq:order_tau}
\end{eqnarray}
In our notation the $Z$-boson interaction with fermions $(f=u,d,e)$ which have electric charge $Q_f$ are given by
\begin{eqnarray}
 {\cal L}_{\rm int}^Z&=&g_Z\bar{f}\gamma^\mu \left[\frac{g_V^f+g_A^f}{2}P_R+\frac{g_V^f-g_A^f}{2}P_L\right]f Z_\mu,\\
g_V^f&=&T^3-2\sin^2(\theta_W)Q_f,\\
g_A^f&=&-T^3,
\end{eqnarray}
where $\theta_W$ is the Weinberg angle and $T^3=1/2$ for up type quarks and $T^3=-1/2$ for down type quarks and charged leptons.
\begin{figure}[t]
 \begin{center}
  \subfigure[]{\includegraphics[width=35mm]{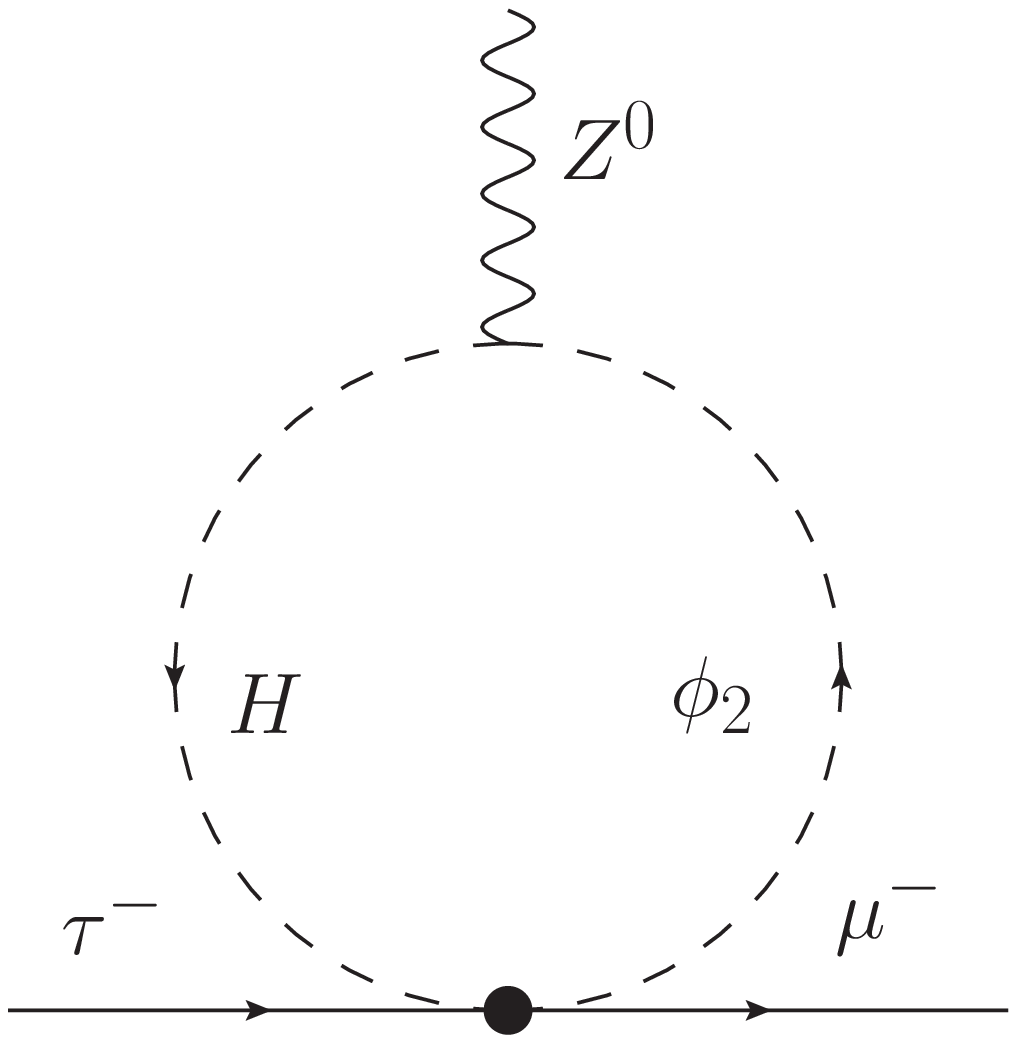}}\qquad
  \subfigure[]{\includegraphics[width=35mm]{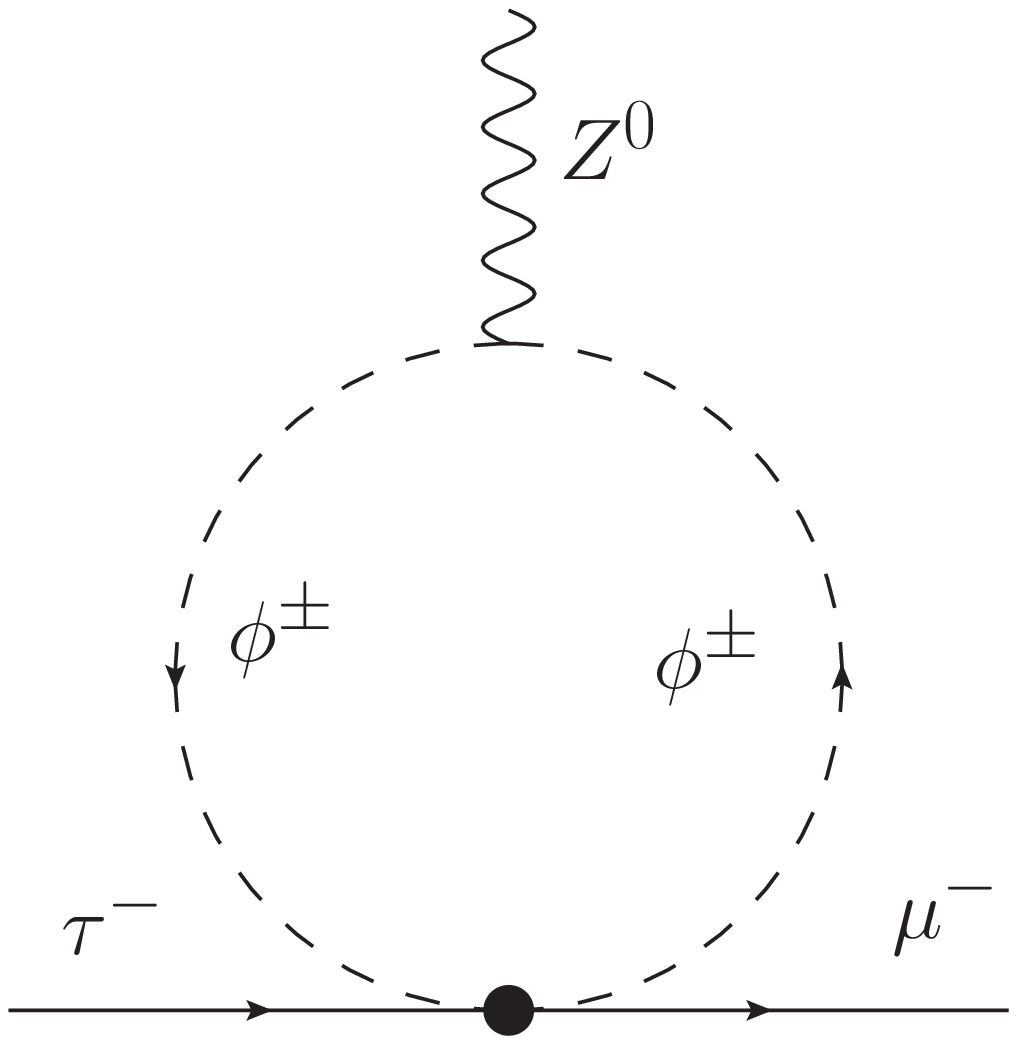}}
\caption{One-loop amplitudes induced by both $\tau\mu h\phi_2$ and $\tau\mu\phi^\mp \phi^\pm$ interactions in Eq.(\ref{eq:effYukawa}) which are denoted as black dots.}  \label{fig:taumuZ_div}
 \end{center}
\end{figure}

Because of the mass relation $m_\tau\ll m_Z,m_H$, the functions $F_V^{v,a}(s)$ and $F_D^{v,a}(s)$ can be approximated by expanding in terms of $r_\tau=m_\tau^2/m_H^2$.
We find $F_V^{v,a}(s)$ and $F_D^{v}(s)$ are of ${\cal O}(r_\tau^0)$ while $F_D^a(s)$ is of ${\cal O}(r_\tau)$.
The fact that $F_V^{v,a}(s)$ and $F_D^{v}(s)$ are the same order stems from the specific features of this model, i.e., the LFV interactions are always accompanied by a chirality flipping.
The leading contribution to the dipole couplings $D_{R,L}^Z(s)$ comes from the diagrams (d) and (e) in Fig.~\ref{fig:tau_muZ}, because these diagrams are not suppressed by $m_\tau$.
Regarding the monopole couplings $C_{R,L}^Z(s)$, every diagram is suppressed by $m_\tau$ because we need another chirality flipping in addition to the LFV couplings.
Therefore, $C_{R,L}^Z(s)$ and $D_{R,L}^Z(s)$ are at the same order in terms of $r_\tau$ in the normalization defined in Eq.~(\ref{eq:Mz}), and thus one cannot neglect $D_{R,L}^Z(s)$ even though the dimension of the dipole operator is higher than that of the monopole one.

The function $F_D^a(s)$ vanishes at the leading order of $r_\tau$ due to the cancellation between the diagrams (d) and (e) of Fig.~\ref{fig:tau_muZ}.
This is again because of the specific chirality structure of this model.
Once we fix the chirality of $\tau^-$ in the external lines, the internal $\tau^-$ and $\mu^-$ in diagrams (d) and (e) have opposite chiralities, which results in the opposite sign of the axial $Z$ boson coupling.

Note that there are other diagrams shown in Fig.~\ref{fig:taumuZ_div} when we consider the full Lagrangian in Eq.~(\ref{eq:effYukawa}).
Although these diagrams are not suppressed by $r_\tau$, they contribute only to the couplings $B_{R,L}^Z(s)$ in Eq.~(\ref{eq:Mz}), and thus we can ignore them.
\subsection{LFV $Z$ decay}\label{sec:3Zdecay}
Using the effective couplings defined in Eq.~(\ref{eq:Mz}), we obtain the following branching fraction of $Z\to\tau\mu$.
\begin{eqnarray}
{\cal B}(Z^0\to \tau^\pm\mu^\mp)&=& {\cal B}(Z^0\to \tau^-\mu^+)+{\cal B}(Z^0\to \tau^+\mu^-)\nonumber\\
&=&\frac{1}{\Gamma_Z}\frac{m_Z}{6\pi}\Bigg[\frac{1}{2}\left(|C_L^Z(m_Z^2)|^2+|C_R^Z(m_Z^2)|^2\right)\nonumber\\
&&+\frac{m_Z^2}{m_\tau^2}\left(\left|D_L^Z(m_Z^2)\right|^2+\left|D_R^Z(m_Z^2)\right|^2\right)\Bigg],\label{eq:zbranch}
\end{eqnarray}
where we neglect contributions with higher order of $r_\tau$.
Here we list the functions $F_V^{v,a}(m_Z^2)$ and $F_D^{v,a}(m_Z^2)$ at the leading order of $r_\tau$:
\begin{eqnarray}
  F_V^a(m_Z^2)&=&-\frac{2\left(4r_Z-1\right)^{3/2}}{r_Z^2}\tan^{-1}\left(\sqrt{4r_Z-1}\right)+8\left(\tan^{-1}{\left(\sqrt{4r_Z-1}\right)}\right)^2+\frac{2}{r_Z^2}{\rm Li}_2(-r_Z)\nonumber\\
&&-\frac{\log{(r_Z)}}{r_Z^2}\left(5r_Z^2-4r_Z-2\log{(r_Z+1)}+1\right)+2\left(\log{(r_Z)}\right)^2\nonumber\\
&&+\frac{21}{2}-\frac{i\pi}{r_Z^2}\left((r_Z^2-2r_Z+2\log{(r_Z+1)}\right),\\
  F_V^v(m_Z^2)&=&\frac{8\sqrt{4r_Z-1}}{r_Z}\tan^{-1}\left(\sqrt{4r_Z-1}\right)-8\left(1-\frac{1}{2r_Z}\right)\left(\tan^{-1}{\left(\sqrt{4r_Z-1}\right)}\right)^2\nonumber\\
&&+\frac{\log{(r_Z)}}{r_Z^2}\left(5r_Z^2-6r_Z+2\log{(r_Z+1)}\right)+\frac{2}{r_Z^2}{\rm Li}_2(-r_Z)\nonumber\\
&&-\frac{1}{2r_Z}\left(17r_Z+(4r_Z+2)\left(\log{(r_Z)}\right)^2-4\right)\nonumber\\
&&-\frac{i\pi}{r_Z^2}\left(r_Z^2-2r_Z+2\log{(r_Z+1)}\right),\\
 F_D^a(m_Z^2)&=&{\cal O}(r_\tau),\\
 F_D^v(m_Z^2)&=&-\frac{4\sqrt{4r_Z-1}}{r_Z}\tan^{-1}\left(\sqrt{4r_Z-1}\right)+\frac{4}{r_Z}\left(\tan^{-1}{\left(\sqrt{4r_Z-1}\right)}\right)^2\nonumber\\
&&+\frac{\log{(r_Z)}}{r_Z}\left(\log{(r_Z)}+2\right)+4,
\end{eqnarray}
where ${\rm Li}_2$ is the dilogarithmic function and $r_Z=m_Z^2/m_H^2$.

We list the numerical values of $F_V^{v,a}(m_Z^2)$ and $F_D^{v,a}(m_Z^2)$ as well as the effective couplings $C_{R,L}^Z(m_Z^2)$ and $D_{R,L}^Z(m_Z^2)$ in TAB.~\ref{tab:ZcoupZ}, where we use the parameters listed in App.~\ref{sec:parameter}.
Except for $F_D^a(m_Z^2)$ which is of ${\cal O}(r_\tau)$, $F$'s are all comparable.
However, the couplings $D_{R,L}^Z(m_Z^2)$ turn out to be two orders of magnitude smaller than $C_{R,L}^Z(m_Z^2)$.
This is caused by a numerical accident in the vector coupling of the $Z$ boson to the charged lepton, $g_V^e=-1/2+2\sin^2{(\theta_W)}=-0.038$. 
By considering the prefactor $m_Z^2/m_\tau^2$ in Eq.~(\ref{eq:zbranch}), the contributions of the dipole and the monopole interactions are comparable.

The imaginary parts of $F$'s are originated from the absorptive part of the diagram in FIG.~\ref{fig:tau_muZc}.
This diagram contributes to $F_V$'s at the leading order of $r_\tau$, while it is not important for $F_D^v$.
For $F_D^a$, since the real part is already ${\cal O}(r_\tau)$, the imaginary part is comparable to the real one.

By using these values the branching ratio is obtained as
\begin{gather}
 {\cal B}(Z^0\to\tau^{\pm}\mu^{\mp})=8.9\times10^{-10}|Y_{\mu\tau}|^2+7.7\times10^{-10}|Y_{\tau\mu}|^2.
\end{gather}
The difference of the coefficients between the first and second terms is caused by parity violation in the weak interaction.
\begin{table}[t]
 \begin{center}

\begin{tabular}[t]{|c| c||c| c|}\hline
$F_V^a(m_Z^2)$ &$5.0-0.78i$ &$C_L^Z(m_Z^2)/Y_{\tau\mu}$ & $(2.3-0.30i)\times10^{-5}$\\ 
$F_V^v(m_Z^2)$ &$-4.8-0.78i$& $C_R^Z(m_Z^2)/Y_{\mu\tau}^*$ &$(-2.0+0.35i)\times10^{-5}$\\
$F_D^a(m_Z^2)$ & $(-8.6+1.6i)\times10^{-5}$&$D_L^Z(m_Z^2)/Y_{\mu\tau}^*$&$-2.7\times10^{-7}$\\
$F_D^v(m_Z^2)$ &$0.84$&$D_R^Z(m_Z^2)/Y_{\tau\mu}$&$-2.7\times10^{-7}$\\ \hline
\end{tabular}

\caption{Coefficients of $g_V^e$ and $g_A^e$ in the effective couplings of dipole and vector operators at $s=m_Z^2$. Regarding the value of $F_D^a(m_Z^2)$ we use the functions defined in Eq.~(\ref{eq:order_tau}).}\label{tab:ZcoupZ}
 \end{center}
\end{table}
\section{LFV tau decays}\label{sec:4taudecay}
In this section, we discuss the LFV tau decays induced by the off-diagonal Yukawa couplings.
We first introduce the general effective Lagrangian responsible for tau decays.
We then calculate the coefficients of the effective operators by using the effective coupling obtained in the previous section.
The branching fractions of the LFV tau decays are evaluated including the $Z$ boson mediated diagrams.
\subsection{Effective four-fermion interactions}
\begin{figure}[t]
 \begin{center}
  \subfigure[]{\includegraphics[width=40mm]{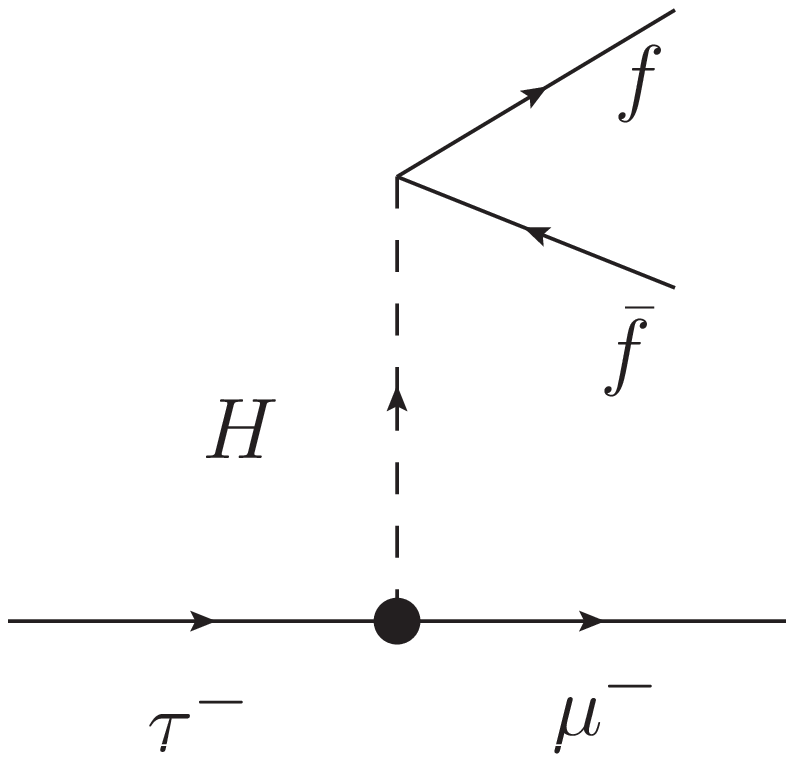}\label{fig:taumuff_a}}
  \subfigure[]{\includegraphics[width=40mm]{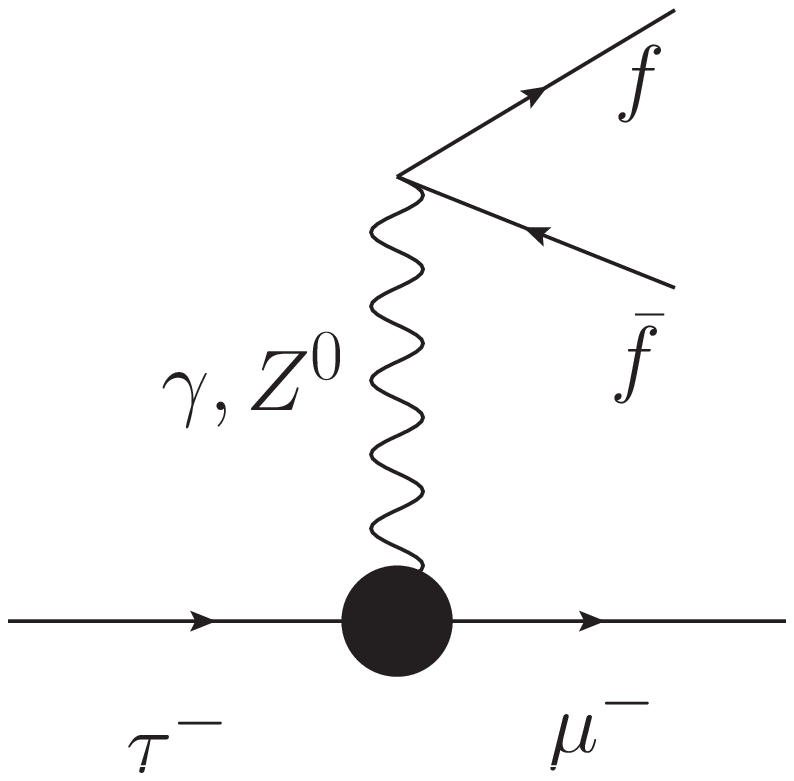}}
   \caption{Diagrams which contribute to effective couplings of four-fermion interactions, where small black dot denotes the LFV Yukawa couplings and large black dot denotes loop induced LFV interactions.}\label{fig:tau_muff}
 \end{center}
\end{figure}

In general, the effective Lagrangian of the processes $\tau\to\mu X, (X=\mu\mu,\rho,\pi,\eta^{(')},a_1)$ is given as follows\footnote{For decays into scalar and pseudoscalar mesons, other operators such as $\tau\mu GG$ and $\tau\mu G\tilde{G}$ can contribute (see Celis {\it et al.}~\cite{Celis:2013xja} for detail).
The operator $\tau\mu G\tilde{G}$ is absent in this model since the Higgs boson do not couple to $G\tilde{G}$.}:
\begin{eqnarray}
{\cal L}_{\rm eff}^{{\rm dim.}6}&=&\frac{D_R^\gamma}{m_\tau} \bar{\tau}_R\sigma^{\mu\nu}\mu_L F_{\mu\nu}+\frac{D_L^\gamma}{m_\tau} \bar{\tau}_L \sigma^{\mu\nu}\mu_R F_{\mu\nu}\nonumber\\
&&+\frac{1}{m_\tau^2}\sum_{f=u,d,s,\mu}\Big\{g_{SLL}^{(f)}(\bar{\tau}_R\mu_L)(\bar{f}_R f_L)+g_{SRR}^{(f)}(\bar{\tau}_L\mu_R)(\bar{f}_L f_R)\nonumber\\
&&+g_{SLR}^{(f)}(\bar{\tau}_R\mu_L)(\bar{f}_Lf_R)+g_{SRL}^{(f)}(\bar{\tau}_L\mu_R)(\bar{f}_Rf_L)\nonumber\\ 
&&+g_{VRR}^{(f)}(\bar{\tau}_R\gamma^\mu\mu_R)(\bar{f}_R\gamma_\mu f_R)+g_{VLL}^{(f)}(\bar{\tau}_L\gamma^\mu\mu_L)(\bar{f}_L\gamma_\mu f_L)\nonumber\\
&&+g_{VRL}^{(f)}(\bar{\tau}_R\gamma^\mu\mu_R)(\bar{f}_L\gamma_\mu f_L)+g_{VLR}^{(f)}(\bar{\tau}_L\gamma^\mu\mu_L)(\bar{f}_R\gamma_\mu f_R)\nonumber\\
&&+g_{TRR}^{(f)}(\bar{\tau}_R\sigma^{\mu\nu}\mu_L)(\bar{f}_R\sigma_{\mu\nu}f_L)+g_{TLL}^{(f)}(\bar{\tau}_L\sigma^{\mu\nu}\mu_R)(\bar{f}_L\sigma_{\mu\nu}f_R)\Big\}+{\rm H.c.}.\label{eq:effL}\\
 {\cal L}_{\rm eff}^{{\rm dim.}7}&=&\frac{1}{m_\tau^3}\sum_{f=u,d,s,\mu}\Big[C_{7LL}^{(f)}(\bar{\tau}_L \sigma^{\mu\nu}\mu_R)\partial_\mu(\bar{f}_L\gamma_\nu f_L)+C_{7LR}^{(f)}(\bar{\tau}_L \sigma^{\mu\nu}\mu_R)\partial_\mu(\bar{f}_R\gamma_\nu f_R)\nonumber\\
&&+C_{7RR}^{(f)}(\bar{\tau}_R \sigma^{\mu\nu}\mu_L)\partial_\mu(\bar{f}_R\gamma_\nu f_R)+C_{7RL}^{(f)}(\bar{\tau}_R \sigma^{\mu\nu}\mu_L)\partial_\mu(\bar{f}_L\gamma_\nu f_L)\Big]+{\rm H.c.},
\end{eqnarray}
where $f$'s are quarks or leptons.
These effective couplings are induced from the diagrams in FIG.~\ref{fig:tau_muff}.
Here we keep the light fermion masses at tree-level while we neglect the contributions of ${\cal O}(m_f/m_\tau)$ in the loop diagrams.
The effective couplings of the photon dipole operators come from one-loop and two-loop diagrams, and are denoted as $D_{R,L}^\gamma=D_{R,L}^{\gamma,{\rm 1loop}}+D_{R,L}^{\gamma,{\rm 2loop}}$.
They are evaluated in Refs.~\cite{Bjorken:1977vt,Chang:1993kw,Leigh:1990kf,Blankenburg:2012ex,Harnik:2012pb}.
The one-loop level contributions are expressed as
\begin{eqnarray}
  D_R^{\gamma,1{\rm loop}}&=&\frac{Q_ee}{2(4\pi)^2}\frac{m_\tau^2}{m_H^2} Y_{\tau\mu}Y_\tau\left(-\frac{4}{3}-\log{\left(\frac{m_\tau^2}{m_H^2}\right)}\right)=-1.0\times10^{-8}Y_{\tau\mu},\\
 D_L^{\gamma,1{\rm loop}}&=&\frac{Q_ee}{2(4\pi)^2}\frac{m_\tau^2}{m_H^2} Y_{\mu\tau}^*Y_\tau\left(-\frac{4}{3}-\log{\left(\frac{m_\tau^2}{m_H^2}\right)}\right)=-1.0\times10^{-8}Y_{\mu\tau}^*,
\end{eqnarray}
where $e$ ($e>0$) is the coupling constant of the electromagnetic interactions.
These couplings are suppressed by three chirality flippings and thus of ${\cal O}(Y_\tau^3)$ considering the normalization defined in Eq.~\eqref{eq:effL}.
On the other hand, there are a class of two-loop Barr-Zee type diagrams~\cite{Barr:1990vd} which are suppressed only by a tau Yukawa coupling, and they are found to be larger than the one-loop ones by about a factor of five, i.e., $D_{R,L}^{\gamma,{\rm 2loop}}=-5.0\times10^{-8}Y_{\tau\mu}(Y_{\mu\tau}^*)$~\cite{Chang:1993kw,Leigh:1990kf,Harnik:2012pb}.

Note here that $D^Z_{R,L}$ calculated in the previous section are of ${\cal O}(Y_\tau)$, and another $m_\tau^2/m_Z^2$ suppression arises when the $Z$ boson is replaced with its propagator.
In total, the $Z$-boson dipole contributions are of ${\cal O}(Y_\tau^3)$, which is the same order as the one-loop photon dipole couplings.
Therefore, we cannot neglect the $Z$-penguin diagrams considering that the dominant two-loop contribution is not much larger than the one-loop one.

The effective couplings of four-fermion operators are listed below.
The coefficients $g_{TRR}^{(f)}$ and $g_{TLL}^{(f)}$ are found to be vanishing in this model.
The scalar couplings $g_S$'s are induced by tree level diagrams which exchange the Higgs boson (FIG.~\ref{fig:taumuff_a}):
\begin{eqnarray}
g_{SRR(L)}^{(f)}&=&\frac{m_\tau^2}{m_H^2} Y_f Y_{\mu\tau}^*,\label{eq:coupH1}\\ 
g_{SLL(R)}^{(f)}&=&\frac{m_\tau^2}{m_H^2} Y_f Y_{\tau\mu},\label{eq:coupH2}
\end{eqnarray}
where we do not neglect the Yukawa coupling of the light fermion.
The vector couplings are sum of the contributions from the photon and the $Z$ boson mediated diagrams as follows:
\begin{eqnarray}
 g_{VHH'}^{(f)}=g_{VHH'}^{(f)}(\gamma)+g_{VHH'}^{(f)}(Z),
\end{eqnarray}
where $H,H'=R,L$, and 
\begin{eqnarray}
g_{VRR(L)}^{(f)}(\gamma)&=&\frac{m_\tau^2}{m_H^2}\frac{\alpha Y_\tau Y_{\mu\tau}^*}{4\pi}Q_e Q_f \left(-\frac{1}{3}\log{\left(\frac{m_\tau^2}{m_H^2}\right)}-\frac{4}{9}\right),\label{eq:vector1}\\
g_{VLL(R)}^{(f)}(\gamma)&=&\frac{m_\tau^2}{m_H^2}\frac{\alpha Y_\tau Y_{\tau\mu}}{4\pi}Q_e Q_f\left(-\frac{1}{3}\log{\left(\frac{m_\tau^2}{m_H^2}\right)}-\frac{4}{9}\right),\label{eq:vector2}\\
g_{VRL}^{(f)}(Z)
&=&\frac{m_\tau^2}{m_Z^2}\frac{\alpha_Z Y_\tau Y_{\mu\tau}^*}{32\pi}(g_V^f-g_A^f)\left[F_V^v(0)g_V^e-F_V^a(0)g_A^e\right],\label{eq:vectorZ1}\\
g_{VRR}^{(f)}(Z)
&=&\frac{m_\tau^2}{m_Z^2}\frac{\alpha_Z Y_\tau Y_{\mu\tau}^*}{32\pi}(g_V^f+g_A^f)\left[F_V^v(0)g_V^e-F_V^a(0)g_A^e\right],\label{eq:vectorZ3}\\
g_{VLR}^{(f)}(Z)
&=&\frac{m_\tau^2}{m_Z^2}\frac{\alpha_Z Y_\tau Y_{\tau\mu}}{32\pi}(g_V^f+g_A^f)\left[F_V^v(0)g_V^e+F_V^a(0)g_A^e\right],\label{eq:vectorZ3}\\
g_{VLL}^{(f)}(Z)
&=&\frac{m_\tau^2}{m_Z^2}\frac{\alpha_Z Y_\tau Y_{\tau\mu}}{32\pi}(g_V^f-g_A^f)\left[F_V^v(0)g_V^e+F_V^a(0)g_A^e\right].\label{eq:vectorZ4}
\end{eqnarray}
The couplings of dimension-seven operators are induced by dipole contribution of $Z$ penguin diagrams as follows:
\begin{eqnarray}
   C_{7LL}^{(f)}
&=&\frac{m_\tau^2}{m_Z^2}\frac{\alpha_Z Y_\tau Y_{\mu\tau}^*}{32\pi}(g_V^f-g_A^f)\left[F_D^v(0)g_V^e+F_D^a(0)g_A^e\right],\label{eq:coup71}\\
 C_{7LR}^{(f)}
&=&\frac{m_\tau^2}{m_Z^2}\frac{\alpha_Z Y_\tau Y_{\mu\tau}^*}{32\pi}(g_V^f+g_A^f)\left[F_D^v(0)g_V^e+F_D^a(0)g_A^e\right],\\
 C_{7RR}^{(f)}
&=&\frac{m_\tau^2}{m_Z^2}\frac{\alpha_Z Y_\tau Y_{\tau\mu}}{32\pi}(g_V^f+g_A^f)\left[F_D^v(0)g_V^e-F_D^a(0)g_A^e\right],\\
 C_{7RL}^{(f)}
&=&\frac{m_\tau^2}{m_Z^2}\frac{\alpha_Z Y_\tau Y_{\tau\mu}}{32\pi}(g_V^f-g_A^f)\left[F_D^v(0)g_V^e-F_D^a(0)g_A^e\right].\label{eq:coup74}
\end{eqnarray}
In the effective couplings of the photon and the $Z$ boson, we take $s=0$, which is a good approximation in the evaluation of tau decays.
The formulae of $F$'s at $s=0$ at the leading order of $r_\tau$ are given by 
\begin{eqnarray}
 F_V^v(0)&=&\frac{6}{1-r_Z}r_Z\log{(r_Z)},\\
 F_V^a(0)&=&\frac{2}{1-r_Z}\left(1-r_Z-2r_Z\log{(r_Z)}\right),\\
 F_D^v(0)&=&-\frac{2r_Z}{(1-r_Z)^2}\left(1+\log{(r_Z)}-r_Z\right),\\
 F_D^a(0)&=&-\frac{r_\tau}{6(1-r_Z)^3}\left(r_Z^3-13r_Z^2+2(r_Z^2+5r_Z-2)\log{(r_Z)}+15r_Z-3\right).
\end{eqnarray}
Their values are listed in TAB.~\ref{tab:Zcoup0}.
The values are not very different from the ones for $s=m_Z^2$ listed in TAB.~\ref{tab:ZcoupZ}.

\begin{table}[t]
 \begin{center}

\begin{tabular}[t]{|c|c||c|c|}\hline
$F_V^a(0)$ &$4.8$ &$C_L^Z(0)/Y_{\tau\mu}$ &$2.2\times10^{-5}$\\ 
$F_V^v(0)$ &$-4.2$& $C_R^Z(0)/Y_{\mu\tau}^*$ &$-1.9\times10^{-5}$\\
$F_D^a(0)$ &$-8.7\times10^{-5}$&$D_L^Z(0)/Y_{\mu\tau}^*$&$-2.5\times10^{-7}$\\
$F_D^v(0)$ &$0.78$&$D_R^Z(0)/Y_{\tau\mu}$&$-2.5\times10^{-7}$\\ \hline
\end{tabular}

\caption{Coefficients of $g_V^e$ and $g_A^e$ in the effective couplings of dipole and vector operators at $s=0$.}\label{tab:Zcoup0}
 \end{center}
\end{table}

\subsection{Branching fractions of the LFV tau decays}
Below, we evaluate the various LFV tau decays which are searched at the B factory experiments.
\subsubsection{$\tau^-\rightarrow\mu^-\mu^+\mu^-$}

The branching ratio of $\tau\to3\mu$ is given by
\begin{eqnarray}
&&  {\cal B}(\tau^-\to\mu^-\mu^+\mu^-)\nonumber\\
&=&\frac{\tau_\tau m_\tau}{128\pi^3}\Big[\frac{2}{9}(-12\log{(\delta)}-13)\left(|eD_R^\gamma|^2+|eD_L^\gamma|^2\right)\nonumber\\
&&+\frac{1}{120}\left(5\left(|C_{7RL}^{(\mu)}|^2+|C_{7LR}^{(\mu)}|^2\right)+4\left(|C_{7RR}^{(\mu)}|^2+|C_{7LL}^{(\mu)}|^2\right)\right)\nonumber\\
&&+\frac{1}{12}\Big(\left|g_{VLR}^{(\mu)}-\frac{1}{2}g_{SLR}^{(\mu)}\right|^2+\left|g_{VRL}^{(\mu)}-\frac{1}{2}g_{SRL}^{(\mu)}\right|^2+2|g_{VRR}^{(\mu)}|^2+2|g_{VLL}^{(\mu)}|^2+\frac{1}{8}(|g_{SLL}^{(\mu)}|^2+|g_{SRR}^{(\mu)}|^2)\Big)\nonumber\\
&&-\frac{1}{6}\Re{\Big[eD_R^\gamma(4C_{7RL}^{(\mu)*}+3C_{7RR}^{(\mu)*})+eD_L^\gamma(4C_{7LR}^{(\mu)*}+3C_{7LL}^{(\mu)*})\Big]}\nonumber\\
&&-\frac{2}{3}\Re{\Big[eD_R^\gamma \left(2g_{VLL}^{(\mu)*}+g_{VLR}^{(\mu)*}-\frac{1}{2}g_{SLR}^{(\mu)*}\right)+eD_L^\gamma \left(2g_{VRR}^{(\mu)*}+g_{VRL}^{(\mu)*}-\frac{1}{2}g_{SRL}^{(\mu)*}\right)\Big]}\nonumber\\
&&-\frac{1}{12}\Re{\Big[2\left(C_{7RL}^{(\mu)*}g_{VLL}^{(\mu)}+C_{7LR}^{(\mu)*}g_{VRR}^{(\mu)}\right)+C_{7RR}^{(\mu)*}\left(g_{VLR}^{(\mu)}-\frac{1}{2}g_{SLR}^{(\mu)}\right)+C_{7LL}^{(\mu)*}\left(g_{VRL}^{(\mu)}-\frac{1}{2}g_{SRL}^{(\mu)}\right)\Big]}
\Big]\nonumber\\
&=&5.5\times10^{-7}|Y_{\mu\tau}|^2+5.5\times10^{-7}|Y_{\tau\mu}|^2,
\end{eqnarray}
where $\tau_\tau$ is the mean life time of the tau lepton.
The cut-off parameter $0<\delta<1$ is introduced to avoid the singularity in the photon mediated contributions.
In the numerical evaluation we assign $\delta=(2m_\mu)^2/m_\tau^2$.

This formula includes the contributions from the dimension-seven operators, which we cannot ignore in general especially when LFV is accompanied by chirality flipping. 
In the model we discuss, however, as we saw in the $Z$ decays, the contribution from the dimension-seven operators is rather suppressed due to small $g_V^e$ and $F_D^a(0)$.

The leading contribution comes from photon dipole operators, and the contributions of four-fermion interactions (mainly the scalar ones) reduce the branching fraction by $9\%$ through the interference terms.
\subsubsection{$\tau^-\to\mu^-\pi^0,\eta,\eta'$}

The four fermion interactions generated by the photon or the Higgs boson exchanges or the photon dipole operator do not contribute to the tau decays into a pseudoscalar meson due to spin and parity.
The leading contribution to such decay modes arises from the effective $Z$ boson couplings.
The branching fraction of $\tau\to\mu\pi$ is given by 
\begin{eqnarray}
 {\cal B}(\tau^-\to\mu^-\pi^0)
&=&\frac{\tau_\tau f_\pi^2}{256\pi m_\tau}\left(1-\frac{m_\pi^2}{m_\tau^2}\right)^2\left[\Big|g_{VRR}^{(u-d)}-g_{VRL}^{(u-d)}\Big|^2+\Big|g_{VLR}^{(u-d)}-g_{VLL}^{(u-d)}\Big|^2\right]\nonumber\\
&=&\tau_\tau\left(\frac{m_\tau^2}{m_Z^2}\right)^2\frac{g_Z^2(g_A^u-g_A^d)^2f_\pi^2}{256\pi m_\tau}\left(1-\frac{m_\pi^2}{m_\tau^2}\right)^2(|C_R^Z|^2+|C_L^Z|^2)\nonumber\\
&=&1.9\times10^{-10}|Y_{\mu\tau}|^2+1.5\times10^{-10}|Y_{\tau\mu}|^2,
\end{eqnarray}
where $g_{VHH'}^{(u-d)}=g_{VHH'}^{(u)}-g_{VHH'}^{(d)}$, $(H,H'=L,R)$, and $f_\pi=130$ MeV is the pion decay constant.
For $\eta$ and $\eta'$, the only difference from $\tau\rightarrow\mu\pi$ is the hadron matrix elements. 
One can obtain the amplitudes of $\tau\to\mu\eta$ and $\tau\to\mu\eta'$ by the replacement of
\begin{eqnarray}
 \frac{g_A^u-g_A^d}{\sqrt{2}}f_\pi\to\frac{g_A^u+g_A^d}{\sqrt{2}}f_P^q+g_A^sf_P^s=g_A^sf_P^s,\qquad (P=\eta,\eta').
\end{eqnarray}
where the decay constants $f_P^q$ and $f_P^s$ are defined following the Feldmann-Kroll-Stech (FKS) mixing scheme~\cite{Feldmann:1998vh}.
We can obtain the branching ratios of $\tau\to\mu\eta$ and $\tau\to\mu\eta'$ as follows:
\begin{eqnarray}
 {\cal B}(\tau^-\to\mu^-P)&=&
\left(\frac{\sqrt{2}g_A^d f_P^s}{f_\pi}\right)^2\left(\frac{m_\tau^2-m_P^2}{m_\tau^2-m_\pi^2}\right)^2{\cal B}(\tau^-\to\mu^-\pi^0),\qquad (P=\eta,\eta').
\end{eqnarray}
From the above relation, we obtain the branching ratios as follows:
\begin{eqnarray}
 {\cal B}(\tau^-\to\mu^-\eta)&=&0.30{\cal B}(\tau^-\to\mu^-\pi^0)\nonumber\\
&=&5.8\times10^{-11}|Y_{\mu\tau}|^2+4.5\times10^{-11}|Y_{\tau\mu}|^2,\\
 {\cal B}(\tau^-\to\mu^-\eta')&=&0.27{\cal B}(\tau^-\to\mu^-\pi^0)\nonumber\\
&=&5.2\times10^{-11}|Y_{\mu\tau}|^2+4.0\times10^{-11}|Y_{\tau\mu}|^2.
\end{eqnarray}
\subsubsection{$\tau^-\to\mu^-\rho^0$}

The spin-parity of the rho meson, $J^P=1^-$, allows the photon-exchange diagram induced from the dipole operator, and the dimension-seven operator and the vector operator induced from $Z$-penguin diagrams as well to contribute to the $\tau\to\mu\rho$ decay.
The branching fraction is given by
\begin{eqnarray}
&& {\cal B}(\tau^-\to\mu^-\rho^0)\nonumber\\
&=&\frac{\tau_\tau f_\rho^2}{32\pi m_\tau}\left(1-\hat{\rho}\right)^2\Big\{\Big(\Big|g_{VRR}^{(u-d)}+g_{VRL}^{(u-d)}\Big|^2\left[\frac{1+2\hat{\rho}}{8}\right]+\Big|\frac{eD_R^{\gamma}}{\hat{\rho}}+\frac{C_{7RR}^{(u-d)}+C_{7RL}^{(u-d)}}{2}\Big|^2\left[\left(4+2\hat{\rho}\right)\hat{\rho}\right]\Big)\nonumber\\
&&+\Re\left[\left(g_{VLL}^{(u-d)}+g_{VLR}^{(u-d)}\right)\left(\frac{eD_R^{\gamma *}}{\hat{\rho}}+\frac{C_{7RR}^{(u-d)*}+C_{7RL}^{(u-d)*}}{2}\right)\right]\left[3\hat{\rho}\right]+(L\leftrightarrow R)\Big\}\nonumber\\
&=&5.5\times10^{-7}|Y_{\mu\tau}|^2+5.8\times10^{-7}|Y_{\tau\mu}|^2,
\end{eqnarray}
where $\hat{\rho}=m_\rho^2/m_\tau^2$, $C_{7HH'}^{(u-d)}=C_{7HH'}^{(u)}-C_{7HH'}^{(d)}$, $(H,H'=L,R)$ and $f_\rho=209$ MeV is decay constant of the rho meson.
Compared to Refs.~\cite{Harnik:2012pb,Celis:2013xja}, we include the contributions from four-fermion and dimension-seven operators.
The leading contribution comes from the dipole operator of the photon and other contributions increase the branching ratio by $5\%$.
\subsubsection{$\tau^-\to\mu^-a_1(1260)$}

As in the case of pseudoscalar modes, axial vector mode is possible only through the $Z$ boson.
The branching ratio is given by
\begin{eqnarray}
&& {\cal B}(\tau^-\to\mu^-a_1(1260))\nonumber\\
&=&\frac{\tau_\tau f_{a_1}^2}{32\pi m_\tau}\left(1-\hat{a}\right)^2\Big\{\Big(\Big|g_{VRL}^{(u-d)}-g_{VRR}^{(u-d)}\Big|^2\left[\frac{1+2\hat{a}}{8}\right]+\Big|\frac{C_{7RR}^{(u-d)}-C_{7RL}^{(u-d)}}{2}\Big|^2\left[(4+2\hat{a})\hat{a}\right]\Big)\nonumber\\
&&+\Re{\left[\left(g_{VLL}^{(u-d)}-g_{VLR}^{(u-d)}\right)\frac{C_{7RR}^{(u-d)*}-C_{7RL}^{(u-d)*}}{2}\right]}\left[3\hat{a}\right]+\left(L\leftrightarrow R\right)\Big\}\nonumber\\
&=&3.5\times10^{-10}|Y_{\mu\tau}|^2+2.5\times10^{-10}|Y_{\tau\mu}|^2,
\end{eqnarray}
where $\hat{a}=m_{a_1}^2/m_\tau^2$ and the decay constant of $a_1$, $f_{a_1}(=230\ {\rm MeV})$, is determined by assuming ${\cal B}(\tau^-\to \nu_\tau a_1^- )={\cal B}(\tau^-\to \nu_\tau 2\pi^-\pi^+ + \nu_\tau 2\pi^0 \pi^- )=18.3\%$~\cite{Schael:2005am,Lee:2010tc}.
\section{Summary}\label{sec:5summary}
After the discovery of the Higgs boson with mass around $125$ GeV, it becomes important to check whether the Higgs boson has properties predicted in the SM.
One important check is to see whether the Higgs boson couples to the mass eigenstates of fermions.
For example, if there is $H\mathchar`-\tau\mathchar`-\mu$ coupling in addition to the SM interactions, non-standard decay of the Higgs boson,  $H\to\tau\mu$, and $Z$ boson, $Z\to\tau\mu$, as well as the various LFV tau decays can happen.

In addition to the photon and the Higgs mediated LFV tau decays studied in Refs.~\cite{Harnik:2012pb,Celis:2013xja}, we complete the analysis by including $Z$ boson mediated contributions.
We calculate one-loop diagrams to generate the effective $\tau\mathchar`-\mu\mathchar`-Z$ interaction and derive formulae as functions of momentum transfer.
We find that at the one-loop level the results are finite and gauge invariant, even though the model corresponds to the addition of a higher dimensional operator to the SM.

In terms of the counting of the $Y_\tau$ insertions, the effective dimension-six and seven four fermion couplings induced from $Z$ penguin diagrams are the same order as one-loop photon penguin diagrams attached to the fermion line.
The contribution of $Z$ penguin diagrams are, however, found to be small, because axial coupling $F_D^a(s)$ is of ${\cal O}(Y_\tau^2)$ and the coefficient of the vector type interaction $g_V^e$ is numerically small in the SM.
The effects of the $Z$ boson couplings are included in the $\tau\to 3\mu$ process, and also we derive the new formulae of the LFV tau decays into pseudoscalar and axial vector mesons in this model.
\section*{Acknowledgments}
This work was supported by JSPS KAKENHI Grant-in-Aid for Scientific Research (B) (No. 15H03669 [RK]), Scientific Research (A) (No. 26247043 [TG]) and MEXT KAKENHI Grant-in-Aid for Scientific Research on Innovative Areas (No. 25105011 [RK,TG]).
\begin{appendices}
\section{Passarino-Veltman functions}\label{sec:PVfunction}
We define the one-loop functions $B_0$ and $C_i$~\cite{Passarino:1978jh,Mertig:1990an}:
\begin{eqnarray}
B_0(p_1^2,m_1^2,m_2^2)&=&(2\pi\mu)^{2\epsilon}\int\frac{d^Dk}{i\pi^2}\frac{1}{N_1 N_2},\\
 \left[C_0,C^\mu,C^{\mu\nu}\right](p_1^2,(p_1-p_2)^2,p_2^2,m_1^2,m_2^2,m_3^2)&=&(2\pi\mu)^{2\epsilon}\int\frac{d^Dk}{i\pi^2}\frac{\left[1,k^\mu,k^\mu k^\nu\right]}{N_1 N_2 N_3},
\end{eqnarray}
where $D=4-2\epsilon$, and
\begin{eqnarray}
 N_1&=&k^2-m_1^2+i\epsilon,\\
 N_2&=&(k+p_1)^2-m_2^2+i\epsilon,\\
 N_3&=&(k+p_2)^2-m_3^2+i\epsilon.
\end{eqnarray}
The tensor integrals can be decomposed by their Lorentz structures as below,
\begin{eqnarray}
 C_{\mu}&=&p_{1,\mu}C_1+p_{2,\mu}C_2,\\
 C_{\mu\nu}&=& g_{\mu\nu}C_{00}+p_{1,\mu}p_{1,\nu}C_{11}+p_{2,\mu}p_{2,\nu}C_{22}+(p_{1,\mu}p_{2,\nu}+p_{2,\mu}p_{1,\nu})C_{12}.
\end{eqnarray}
The explicit expression of these tensor functions are
\begin{eqnarray}
B_0(a,b_1,b_2)&=&\int_{0}^{1}dx\left[-\log{\left(\Delta_B(x)\right)}+\Lambda+2\log{\mu}\right],\\
 C_0(a_1,a_2,a_3,b_1,b_2,b_3)&=&\int^1_0dx\int^{1-x}_0dy\frac{-1}{\Delta_C(x,y)},\\
 C_1(a_1,a_2,a_3,b_1,b_2,b_3)&=&\int^1_0dx\int^{1-x}_0dy\frac{y}{\Delta_C(x,y)},\\
 C_2(a_1,a_2,a_3,b_1,b_2,b_3)&=&\int^1_0dx\int^{1-x}_0dy\frac{1-x-y}{\Delta_C(x,y)},\\
C_{00}(a_1,a_2,a_3,b_1,b_2,b_3)&=&\int^1_0dx\int^{1-x}_0dy\Big[-\frac{1}{2}\log{\left(\Delta_C(x,y)\right)}+\frac{1}{2}\Lambda+\log{\mu}\Big],\\
C_{11}(a_1,a_2,a_3,b_1,b_2,b_3)&=&-\int^1_0dx\int^{1-x}_0dy\frac{y^2}{\Delta_C(x,y)},\\
C_{22}(a_1,a_2,a_3,b_1,b_2,b_3)&=&-\int^1_0dx\int^{1-x}_0dy\frac{(1-x-y)^2}{\Delta_C(x,y)},\\
C_{12}(a_1,a_2,a_3,b_1,b_2,b_3)&=&-\int^1_0dx\int^{1-x}_0dy\frac{y(1-x-y)}{\Delta_C(x,y)},
\end{eqnarray}
where $\Lambda=1/\epsilon -\gamma +\log{(4\pi)}$, $\gamma$ is the Euler constant and
\begin{eqnarray}
 \Delta_B(x)&=&(b_1-(1-x)a)x+(1-x)b_2,\\
\Delta_C(x,y)&=&-a_1 xy-(a_2 y+a_3 x)(1-x-y)+b_1x+b_2y+b_3(1-x-y).
\end{eqnarray}
\section{Input parameters}\label{sec:parameter}
The input parameters used in numerical evaluations are listed in TAB.~\ref{tab:param}.
\begin{center}

\begin{tabular}[t]{|c|c||c|c|} \hline
$\alpha$ &$128^{-1}$&$\Gamma_Z$ & $2.49$ GeV \\ 
$g_Z$ & $0.741$&$f_\pi$ & $0.130$ GeV \\
$\sin^2{\theta_W}$ & $0.231$ & $f_{\eta}^s$ & $-0.110$ GeV \\ 
$v$ & $246$ GeV & $f_{\eta'}^s$ & $0.135$ GeV \\
$m_H$ & $125$ GeV & $f_\eta^q$ & $0.108$ GeV \\ 
$\tau_\tau$ & $4.41\times10^{11}$ ${\rm GeV}^{-1}$ & $f_{\eta'}^q$ & $0.088$ GeV \\ 
$m_\tau$ & $1.78$ GeV & $f_\rho$ & $0.209$ GeV \\ 
$m_\mu$ & $0.106$ GeV & $f_{a_1}$ & $0.230$ GeV \\
$m_Z$ & $91.2$ GeV & $m_\eta$ & $0.548$ GeV \\ 
$m_\rho$ & $0.770$ GeV & $m_{\eta'}$ & $0.958$ GeV \\
$m_{a_1}$ & $1.23$ GeV & $m_\pi$ & $0.140$ GeV  \\ \hline
\end{tabular} 
\begin{table}[h]
 \caption{Input parameters.}\label{tab:param}
\end{table}
\end{center}

The decay constants of isospin-triplet hadrons are defined as
\begin{eqnarray}
 -i\sqrt{2}f_\pi p^{\mu}&=&\Braket{0|\left(\bar{u}\gamma^\mu \gamma^5 u-\bar{d}\gamma^\mu \gamma^5 d\right)|\pi(p)},\\
 -i\sqrt{2}m_{a_1}f_{a_1} \epsilon_{a_1}^{\mu}(p)&=&\Braket{0|\left(\bar{u}\gamma^\mu \gamma^5 u-\bar{d}\gamma^\mu \gamma^5 d\right)|a_1(p)},\\
 -i\sqrt{2}m_{\rho}f_\rho \epsilon_{\rho}^{\mu}(p)&=&\Braket{0|\left(\bar{u}\gamma^\mu  u-\bar{d}\gamma^\mu d\right)|\rho(p)},
\end{eqnarray}
where $p^\mu$ is a four momentum of hadrons.
The decay constants of $\eta$ and $\eta'$, $f_{\eta^{(\prime)}}^{q,s}$, are defined as 
\begin{eqnarray}
-i\sqrt{2}f_P^q p^\mu &=& \Braket{0|\left(\bar{u}\gamma^\mu \gamma^5 u+\bar{d}\gamma^\mu \gamma^5 d\right)|P(p)},\\
-if_P^s p^\mu &=& \Braket{0|\bar{s}\gamma^\mu \gamma^5 s|P(p)},
\end{eqnarray}
where $P=\eta,\ \eta'$.
The values in TAB.~\ref{tab:param} are given in Ref.~\cite{Beneke:2002jn}.
\end{appendices}
\bibliographystyle{h-physrev3.bst}
\bibliography{ref.bib}
\end{document}